\documentclass[lettersize,journal]{IEEEtran}
\usepackage{amsmath,amsfonts}
\usepackage[linesnumbered,ruled,lined]{algorithm2e}
\usepackage{array}
\usepackage{subfigure}
\usepackage{textcomp}
\usepackage{stfloats}
\usepackage{url}
\usepackage{verbatim}
\usepackage{graphicx}
\usepackage{cite}
\usepackage{bm}
\usepackage{booktabs}
\usepackage{tcolorbox}
\usepackage{soul} 
\begin{document}

\title{Anti-jamming Method for SAR Using \\Joint Waveform Modulation and Azimuth Mismatched Filtering}

\author{Zhuan Sun,~\IEEEmembership{Student Member,~IEEE,} and Zhanyu Zhu,~\IEEEmembership{Member,~IEEE}
    \thanks{This work was supported in part by the National Natural Science Foundation of China under Grant U2130202. (\textit{Corresponding author: Zhanyu Zhu} ). }   
	\thanks{The authors are with the School of Electronic and Information Engineering, Soochow University, Suzhou, 215006, China (e-mail: zsun1@stu.suda.edu.cn; e-mail: zyzhu@suda.edu.cn).}
}

\markboth{IEEE GEOSCIENCE AND REMOTE SENSING LETTERS}%
{Shell \MakeLowercase{\textit{et al.}}: Anti-jamming Method for SAR Using Joint Waveform Modulation and Azimuth Mismatched Filtering}

\maketitle

\begin{abstract}
High-fidelity deception jamming can seriously mislead Synthetic Aperture Radar (SAR) image interpretation and target detection, which is difficult to identify or eliminate through traditional anti-jamming methods. Based on the Range-Doppler Algorithm (RDA), an anti-jamming method for SAR by using joint waveform modulation and azimuth mismatched filtering is proposed in this paper. The signal model of SAR echo after pulse compression with deception jamming is derived from the radar detection and jamming characteristics. A multi-objective cost function is introduced to suppress the jamming level and keep the real target information, parameterized with the waveform initial phase and the azimuth mismatched filter coefficients, which is optimized using the second-order Taylor expansion approximation and the alternating direction multiplier method (ADMM). The performance of this proposed method is evaluated through the convergence analysis and anti-jamming experiments on the point target and the distributed target scenarios.
\end{abstract}


\section{Introduction}
\IEEEPARstart{S}{ynthetic} Aperture Radar (SAR) is widely used in military and civil fields, such as terrain mapping, ocean observation, disaster prediction and so on \cite{Strozzi2006terrain,Makhoul2015GMTI,Kundu2022Flood}. However, radar systems that work at high altitude have the risk of being interfered by electromagnetic (EM) signals in real  environments. In particular, the existence of active jamming methods, such as deception jamming, has become a major threat to radar systems, which is difficult to identify or eliminate by traditional anti-jamming methods \cite{Wen2019Cognitive}. Jammers often use digital radio frequency memory (DRFM) technology to modulate and forward intercepted radar signals to produce realistic false targets \cite{Feng2017JammingWideband}. The deception jamming signal generated is basically the same as the radar target signal, and a variety of false information is added to modulate it, which will directly affect the perception performance of radar equipments.

To the best of our knowledge, the existing research can be basically divided into two categories, namely, the receiving processing method and the transmission modulation method. The former method, such as the multi-channel method \cite{Rosenberg2006multichannel} or dynamic synthetic aperture method \cite{Zhao2017Single}, has the disadvantages of high cost of deployment and maintenance, incomplete separation of high-fidelity jamming, and residual energy. The latter method reduces the probability of interception and jamming through the temporal varying signal design. In \cite{Wu2018Nonperiodic}, a nonperiodic interrupted sampling-linear frequency modulation (NIS-LFM) radar transmitting signal has been reported for anti-jamming, while the energy loss of NIS-LFM signal needs to be compensated to maintain range profile. The azimuth phase coding (APC) method by shifting the jammed signal to suppress deceptive jamming was proposed by Tang et. al., but can not fully tackle the problem of residual jamming energy \cite{Tang2021HighFidelity}. Zhou et. al. proposed a method to counter the interrupted-sampling repeater jamming (ISRJ) by transmitting a phase-coded (PC) waveform and designing the corresponding mismatched filter. However, the waveform and filter generated by this algorithm may reduce the imaging performance in the range direction \cite{Zhou2022Waveform}.

The matched filter is always used to demodulate the received signal, called pulse compression, to maximize the signal-to-noise ratio (SNR) of the received signal, but cannot suppress the coherent jamming and may produce a high level of sidelobe. Then, a range mismatched filtering technique was developed to effectively suppress sidelobe energy. Range mismatched filtering and waveform design, aiming at range signals, can deal with periodic jamming such as unintentional jamming, ISRJ, etc \cite{Levanon2009Range, Liu2018RangeMismatchedfilter,Zhou2022Waveform}. The periodic jamming form is simply repeated in each pulses, whose characteristics just vary during pulse duration. However, the deception jamming we aim at has the inter-pulse varying characteristic, which can produce faked targets on the image. Since the range mismatched filter is determined, it can not properly deal with the varied jamming signal in different pulses, named space-variant jamming.
According to the imaging theory of the Range-Doppler Algorithm (RDA) \cite{Bao2005imaging}, SAR imaging processing can be approximately separated into two matched filtering processes in range and azimuth dimensions, respectively. Based on RDA imaging theory, a method of SAR anti-jamming using multi-objective function to optimize the initial phase of the transmitted waveform and the coefficients of azimuth mismatched filtering is proposed in this paper. It can handle space-variant jamming and suppress the deception jamming level without losing the performance of the pulse compression in range direction.

The remainder of this paper is organized as follows. Section II describes the signal model of the transmitted signal and jamming signal. Section III presents the method of joint waveform modulation and azimuth mismatched filtering to suppress the deception jamming and keep the imaging performance. Section IV provides convergence analysis and experimental cases of point target imaging and distributed target imaging. Section V summarizes the work.

\section{Signal Model}
Imaging radar generally adopts a linear frequency modulation (LFM) signal as the transmitted signal, which is received by the radar system with interaction of the target. At the receiving end, the analog echo signal is converted down and sampled into a baseband digital signal. Without losing generality, the transmitted signal selected in this paper is the LFM phase coded signal of inter-pulse diversity, and the baseband signal can be expressed as \cite{Bao2005imaging}
\begin{align}\label{equation1}
		s \left({{t}_{\text{r}}},{{t}_{\text{a}}}\right)
		=\text{rect}\left(\frac{{t}_{\text{r}}}{{T}_{\text{r}}}\right)\text{exp} \left(\text{j}
		\pi{{K}_{\text{r}}}
		t_{\text{r}}^{\text{2}}
		+\text{j}{{\varphi}_{n}} \right)
\end{align}
where ${{t}_{\text{r}}}$ is the fast time in range direction, ${{t}_{\text{a}}}$ is the slow time in azimuth direction, ${{{K}_{\text{r}}}={{{B}_{\text{r}}}}/{{{T}_{\text{r}}}}}$ is the frequency modulation slope with ${B}_{\text{r}}$ and ${T}_{\text{r}}$ being the signal bandwidth and pulse duration, respectively, ${{\varphi}_{n}}$ is the encoding phase of the current pulse repetition cycle, and ${\text{rect}\left( {{{t}_{\text{r}}}}/{{{T}_{\text{r}}}} \right)}$ is the rectangular window function. Since ${{\varphi}_{n}}$ varies with ${{t}_{\text{a}}}$, the transmitted signal is space-variant. The echo signal is captured from the scattering electronic field of the transmitted signal by the target. After the down conversion frequency mixing, the carrier frequency is eliminated and the baseband signal can be written as
\begin{align}\label{equation2}
	s\left({{t}_{\text{r}}},{{t}_{\text{a}}}\right)=
	&\sigma \cdot \text{rect}\left( \frac{{{t}_{\text{r}}}-\tau }{{{T}_{\text{r}}}} \right)
	\cdot \text{rect}\left( \frac{{{t}_{\text{a}}}}{{{T}_{\text{a}}}} \right)
	\cdot \text{exp} \left( \text{j}\varphi \right) \notag\\
	&\cdot \exp \left( -\text{j2}\pi {{f}_{\text{c}}}\tau \right)
	\cdot \exp \left[ \text{j}\pi {{K}_{\text{r}}}{{\left( {{t}_{\text{r}}}-\tau \right)}^{2}} \right]
\end{align}
where $\tau ={2{{R}_{\text{T}}}\left({{t}_{\text{a}}}\right)}/{\text{c}}$ with $R$ and $\text{c}$ being signal propagation distance and the light velocity, respectively, and $\varphi ={{\varphi }_{n}}$, if the detected target is a real target without jamming. Similarly, if the signal comes from a faked target with jamming, then ${\tau={2{{R}_{\text{J}}}\left({{t}_{\text{a}}}\right)}/{\text{c}}}$ and $\varphi={{\varphi }_{m}}$, where ${{\varphi }_{m}}$ is the phase corresponding to the pulse cycle identified by the jammer, so jamming signal is also a space-variant signal. Besides, ${{f}_{\text{c}}}$ is the carrier frequency and $\exp \left( -\text{j2 }\!\!\pi\!\!\text{ }{{f}_{\text{c}}}\tau \right)$ is the Doppler shift. Then, the signal is convolved with the filtering function ${h\left({{t}_{\text{r}}}\right)=\text{exp}\left(- \text{j}\pi{{K}_{\text{r}}}t_{\text{r}}^{\text{2}} \right)}$ to perform the range matched filtering. Thus, the compressed signal can be written as
\begin{align}\label{equation3}
 {{s}_{\text{T}}}\left({{t}_{\text{r}}},{{t}_{\text{a}}}\right)
	=& \sigma \cdot {{T}_{\text{r}}}
	\cdot \text{rect}\left( \frac{{{t}_{\text{a}}}}{{{T}_{\text{a}}}} \right)
	\cdot \exp \left( j\varphi-\text{j2}\pi {{f}_{\text{c}}}\tau \right) \notag\\
	& \cdot \exp \left[ \text{j}\pi {{K}_{\text{r}}}\left( {{\tau }^{2}}-{{t}_{\text{r}}^{2}} \right) \right]
	 \cdot \exp \left[ \text{j}\pi {{K}_{\text{r}}} {{\left( {{t}_{\text{r}}}-\tau \right)}^{2}} \right] \notag\\
	&\cdot \text{sinc}\left[ \text{j}\pi {{K}_{\text{r}}} \left( {{t}_{\text{r}}}-\tau \right){{T}_{\text{r}}} \right].
\end{align}

Since $R\left({{t}_{a}}\right)$ in \eqref{equation3} varies with ${t}_{a}$, range cell migration correction (RCMC) is required when the migration distance is greater than the range resolution cell \cite{Bao2005imaging}. After compensation, the position of the waveform envelope and the high-order phase error can be ignored due to their slight impact on the imaging processing. It is worth noting that, since the deception target is faked by the jammer, it has a fixed Doppler shift in each echo signal. The deception signal can not be fully compensated for the migration difference from the real target. Therefore, the phase and time-delay items related to the transport delay are retained in the derived equation.

According to the difference between the real target signal and the jamming signal mentioned above, the signal sequence after pulse compression and RCMC of a point target can be presented as ${{\bm{s}}=[{{s}_{\text{1}}},{{s}_{\text{2}}},\cdots ,{{s}_{N}}]^{\text{T}}}$, and the mismatched filtering coefficient in azimuth direction is ${{\bm{h}}=[{{h}_{\text{1}}},{{h}_{\text{2}}},\cdots ,{{h}_{N}}]^{\text{T}}}$, where ${h}_{\text{i}} = A_\text{i} \text{exp}\left(- \text{j}2\pi{{V}^{2}} t_{\text{a}}^{2}/\left({\lambda R_\text{c}} \right) + \theta_\text{i} \right)$, $V$ represents the speed of the radar platform, $R_\text{c}$ represents the slant range from the radar to the scene center with $V$ and $R_\text{c}$ being fixed values, ${h}_{\text{i}}$ is derived from the azimuth processing of RDA, and $A_\text{i}$ and $\theta_\text{i}$ are variables to be optimized. Then, scene imaging is obtained by convolving ${\bm{h}}$ with \eqref{equation3} in $t_{\text{a}}$ dimension, namely azimuth direction. The extension matrix formation of ${\bm{S}}$ and ${\bm{H}}$ can be expressed \cite{Zhou2022Waveform}, respectively, as 
\begin{align}
	{\bm{S}}={{\left[ \begin{matrix}
				{{s}_{N}} & {{s}_{N-1}} & \cdots  & {{s}_{1}} & 0 & \cdots  & 0  \\
				0 & {{s}_{N}} & \cdots  & {{s}_{2}} & {{s}_{1}} & \cdots  & 0  \\
				\vdots  & \vdots  & \ddots  & \vdots  & \vdots  & \ddots  & \vdots   \\
				0 & 0 & \cdots  & {{s}_{N}} & {{s}_{N-1}} & \cdots  & {{s}_{1}}  \\
			\end{matrix} \right]}^{\text{H}}}
\end{align}
and
\begin{align}
	{\bm{H}}={{\left[ \begin{matrix}
				0 & \cdots  & 0 & {{h}_{\text{1}}} & \cdots  & {{h}_{N\text{-1}}} & {{h}_{N}}  \\
				0 & \cdots  & {{h}_{\text{1}}} & {{h}_{\text{2}}} & \cdots  & {{h}_{N}} & 0  \\
				\vdots  &  & \vdots  & \vdots  &  & \vdots  & \vdots  \\
				{{h}_{\text{1}}} & \cdots  & {{h}_{N\text{-1}}} & {{h}_{N}} & 0 & \cdots  & 0  \\
			\end{matrix} \right]}^{\text{H}}}
\end{align}
where
\begin{align} \label{equation6}
	{{s}_{i}}=\text{exp}\left[ -{\text{j4}\pi R}/{\lambda}+\text{j}\pi{{K}_{\text{r}}}{{\left( {2R}/{\text{c}} \right)}^{2}}\text{+j}{{\varphi}_{i}} \right].
\end{align}

The deception jamming signal in matrix form can be expressed as ${{\bm{s}}_{\text{jam}}}={{\bm{Js}}}$ with ${\bm{\xi}}={{[ {{\xi}_{\text{1}}},{{\xi}_{\text{2}}},\cdots ,{{\xi}_{N}} ] }^{\text{T}}}$ and ${{\bm{J}}}=\text{diag}\left\{ {\bm{\xi}} \right\}$ being the response vector of the jammer, where ${{\xi}_{i}}={{\sigma }_{\text{Ji}}}\cdot \exp \left(-\text{j4}\pi{\Delta {{R}_{\text{J}i}}}/{\lambda }+j {{\varphi }_{m-n}}\right)$ and the corresponding extended matrix is similar to ${\bm{S}}$ of the matrix formation, which is represented as ${\bm{\mathit{\Psi}}}$ .

\section{The Joint Waveform Modulation and Azimuth Mismatched Filering Method}
To ensure the imaging quality and anti-jamming performance of the detected target, the cost function is designed as follows:
\begin{align}\label{equation7}
	 \underset{\bm{s},\bm{h}}{\mathop{\text{min}}}\,
	 f\left(\bm{s},\bm{h}\right)
		=&{\alpha }_{\text{1}}
		{{\bm{h}}^{\text{H}}} {\bm{S}^{\text{H}}}{\bm{D S h}}
		+{{\alpha }_{\text{2}}}
		{{\bm{h}}^{\text{H}}} {{\bm{\mathit{\Psi}}}^{\text{H}}} {\bm{\mathit{\Psi} h}} \notag\\
    	& +{{\alpha }_{3}}{{\left| {\bm{d S h}} - {{\beta }_{1}} {{\bm{s}}^{\text{H}}} {\bm{h}} \right|}^{2}}
			+{{\alpha }_{4}}{{\left| {{\bm{h}}^{\text{H}}}{\bm{s}}-{{\beta }_{2}} \right|}^{2}} \\
		\text{s.t.} \  {{\bm{h}}^{\text{H}}}{\bm{h}} = N &\ ,  \left| {{s}_{n}} \right|=1,n=1,2,\ldots ,N	\notag
\end{align}
where ${\alpha }_{\text{1}}$, ${{\alpha }_{\text{2}}}$, ${{\alpha }_{3}}$ and ${{\alpha }_{4}}$ are the weights of constraint, four terms in the right-hand side of $f\left({\bm{s}},{\bm{h}}\right)$ represent the sidelobe energy integral of target signal, the energy integral of deception jamming signal, the integral side lobe ratio (ISLR) constraint, and the loss process gain (LPG) of the mismatched filtering constraint, respectively. Moreover, ${\bm{D}}=\text{diag}\left\{ {\bm{d}} \right\}$, where ${\bm{d}}$ represents the signal sidelobe. Note ${\bm{d}}$ is a row vector of length $2N-1$ with the $N$-th element zero. ${{\beta }_{1}}$ is the predefined RISL value with ${{\beta}_{1}}={{10}^{{{R}_{\text{ISL}}}/10}}$, ${\beta}_{2}$ is the predefined value of the filtering gain, wherein the expressions of ISLR and LPG can be found in formulas (8) and (9) in literature \cite{Pishrow2021LPG_ISRL}.

With the energy constraint, the solution of the cost function becomes a two-variable quadratic optimization problem, which is difficult to solve directly. Therefore, the Alternating Direction Multiplier Method (ADMM) was introduced to optimize the function with the constant term ignored without affecting the result. Firstly, it is assumed that the transmitted signal $\bm{s}$ is a priori knowledge, and the mismatched filter coefficient $\bm{h}$ can be optimized by the following model:
\begin{align}\label{equation8}
		{\bm{h}^{(k+\text{1})}}
		& =\underset{\bm{h}}{\mathop{\arg \min }}\,f\left({{\bm{s}}^{(k)}},{\bm{h}}\right) \notag\\
		& =\underset{\bm{h}}{\mathop{\arg \min }}\,{{\bm{h}}^{\text{H}}}{{\bm{Z}}^{(k)}}{{\bm{h}}}+{{\bm{h}}^{\text{H}}}{{\bm{z}}^{(k)}} \\
		\text{s.t.} \ {{\bm{h}}^{\text{H}}} & {\bm{h}}= N  	\notag
\end{align}
where
\begin{align}
	\label{equation9}
	{\bm{Z}}
		= & {\alpha }_{\text{1}}{{\bm{S}}^{\text{H}}}\bm{DS}+{{\alpha }_{\text{2}}}{{\bm{\mathit{\Psi} }}^{\text{H}}}{\bm{\mathit{\Psi}}}
		-2{{\alpha}_{3}}{{\beta }_{1}}\operatorname{Re}\left(\bm{sdS}\right)\notag\\
		& +{{\alpha}_{3}}{{\bm{S}}^{\text{H}}}{{\bm{d}}^{\text{H}}} \bm{dS}
		+{{\alpha}_{3}}\left({{\beta }_{1}}^{2}+{{\alpha }_{4}}\right)
		\bm{s} {{\bm{s}}^{\text{H}}}  \\
	\label{equation10}	
	\bm{z}=&-2{{\alpha }_{4}}{{\beta }_{2}}\bm{s}.
\end{align}

To avoid operations of high computational complexity such as reverse operation, the current cost function is converted into the second order Taylor expansion approximation \cite{Jyothi2019MajorizationMinimization}, which can be written as
\begin{align}\label{equation11}	
	{g_{\text{1}}} \left({{\bm{s}}^{(k)}},\bm{h} | {{\bm{h} }^{(k)}} \right)
	=& f\left({{\bm{s}}^{(k)}},{{\bm{h} }^{(k)}}\right) \notag\\
	& +\nabla { f{\left({{\bm{s}}^{(k)}},{{\bm{h} }^{(k)}}\right)}^{\text{H}}}\left(\bm{h} -{{\bm{h} }^{(k)}}\right)\notag\\
	& +\frac{1}{2}{{\left(\bm{h} -{{\bm{h} }^{(k)}}\right)}^{\text{H}}}{{\bm{\Lambda}}_{1}}^{(k)}
	\left(\bm{h} -{{\bm{h}}^{(k)}}\right) \notag\\
	& \ge f\left({{\bm{s}}^{(k)}},\bm{h} \right)	
\end{align}
where
\begin{align} \label{equation12}
	{{\bm{\Lambda} }_{1}}^{(k)}=\left(\underset{i} {\mathop{\max}}\,\sum\limits_{j=1}^{N}
	 {{{\nabla }^{2}}f\left({{\bm{s}}^{(k)}},{{\bm{h} }^{(k)}}\right)}\right){{\bm{I}}_{N}} \\
	 \label{equation13}
	{\nabla { f{\left({{\bm{s}}^{(k)}},{{\bm{h} }^{(k)}}\right)}}}
		= 2\left(\bm{Zh}+\bm{z}\right),
	{{{\nabla }^{2}}f\left({{\bm{s}}^{(k)}},{{\bm{h} }^{(k)}}\right)} = \bm{Z}
\end{align}
Then, $f{\left({{\bm{s}}^{(k)}},{{\bm{h} }^{(k)}}\right)}$ in \eqref{equation8} is replaced by ${g_{\text{1}}} \left({{\bm{s}}^{(k)}},\bm{h} | {{\bm{h} }^{(k)}} \right)$ in \eqref{equation11}. Using the method of Lagrange multiplier in \eqref{equation8} leads to
\begin{align} \label{equation14}
	{{\bm{h}}^{(k+1)}}=\frac{ \sqrt{N}{\left[-4\left({{Z}^{(k)}}{{\bm{h} }^{(k)}}+{{z}^{(k)}}\right)+{{\bm{\Lambda}}_{1}}^{(k)}{{\bm{h} }^{(k)}} \right] } }
	{\sqrt{\left\| -4\left({{Z}^{(k)}}{{\bm{h} }^{(k)}}+{{z}^{(k)}}\right)+{{\bm{ \Lambda}}_{1}}^{(k)}{{\bm{h} }^{(k)}} \right\|_{2}^{2}}}.
\end{align}	

To facilitate the solution of variable $\bm{s}$, \eqref{equation7} is equivalent to
\begin{align}\label{equation15}	
		\underset{\bm{s},\bm{h} }{\mathop{\text{min}}}\,f\left(\bm{s},\bm{h} \right)
		= & {\alpha }_{\text{1}}{{\bm{s}}^{\text{H}}}{{\bm{H} }^{\text{H}}}\bm{DHs} +{{\alpha }_{\text{2}}}{{\bm{s} }^{\text{H}}} \bm{J} {{\bm{H}}^{\text{H}}}\bm{HJs} \notag\\
		&+{{\alpha }_{3}}{{\left| \bm{dHs}-{{\beta }_{1}}{{\bm{h} }^{\text{H}}}\bm{s} \right|}^{2}}+{{\alpha }_{4}}{{\left| {{\bm{s} }^{\text{H}}}\bm{h} -{{\beta }_{2}} \right|}^{2}} \\
		\text{s.t.} \ {{{\bm{h}}}^{\text{H}}}{{\bm{h}}}=N &\ , \left| {{s}_{n}} \right|=1,n=1,2,\ldots ,N  	 \notag
\end{align}
where $\bm{J}$ is represented by the state at $k$ because the variable $\bm{S}$ changes slowly in the iterative solution process. Then, by assuming that the coefficient of the mismatched filter $\bm{h}$ is known, the transmitted signal $\bm{s}$ can be optimized by the following equation:
\begin{align} \label{equation16}
		{{\bm{s}}^{(k+\text{1})}}
		=& \underset{\bm{s} }{\mathop{\arg \min }}\,f\left(\bm{s},{{{\bm{h}}}^{(k+\text{1})}}\right)\notag\\
		=& \underset{\bm{s}}{\mathop{\arg \min }}\,{{\bm{s}}^{\text{H}}}{{\bm{Y}}^{(k+\text{1})}}\bm{s} +{{\bm{s} }^{\text{H}}}{{\bm{y}}^{(k+\text{1})}} \\ 		
		\text{s.t.} \ \  \left| {{s}_{n}} \right| &  =1  \ ,n=1,2,\ldots ,N 	\notag
\end{align}
where
\begin{align}	
	\label{equation17}
	\bm{Y}
		=& {\alpha }_{\text{1}}{{\bm{H} }^{\text{H}}}\bm{DH}+{{\alpha }_{\text{2}}}\bm{J}{{\bm{H}}^{\text{H}}}\bm{HJ}
		-2{{\alpha }_{3}}{{\beta }_{1}}\operatorname{Re}\left(\bm{hdH}\right) \notag\\
		& +{{\alpha }_{3}}{\bm{h H}^{\text{H}}}{{\bm{d}}^{\text{H}}}\bm{ dH}
		+{{\alpha }_{3}}\left( {{\beta }_{1}}^{2}+{{\alpha }_{4}} \right) {\bm{h}}{{{\bm{h}}}^{\text{H}}}	
		\\
    \label{equation18}
	\bm{y}=& -2{{\alpha }_{4}}{{\beta }_{2}}{\bm{h}}.
\end{align}
Using the same solving method of \eqref{equation8} to settle the problem of \eqref{equation16}, we have
\begin{align}\label{equation19}
	{{\bm{s}}^{(k+1)}}=\exp\left(\text{j}
	\arg \left[ \left(-4{{\bm{Y}}^{(k+1)}}+{{\bm{\Lambda }}_{2}}^{(k+1)}\right)
	{{\bm{s}}^{(k)}}-4{{\bm{y}}^{(k+1)}}\right]\right)
\end{align}
where
\begin{align}\label{equation20}
	{{\bm{\Lambda} }_{2}}^{(k+1)}=\left(\underset{i} {\mathop{\max}}\,\sum\limits_{j=1}^{N}
	{2{{\bm{Y}}^{(k+1)}}}\right){{\bm{I}}_{N}}.
\end{align}
Then, the value of the initial phase ${\varphi}$ can be  determined according \eqref{equation6}.

\begin{algorithm}[!t] 	\label{alg1}
\caption{WM-AMMFA}
		{\bf{Initialize}}:$ {{s}^{(0)}},\  {{h}^{(0)}},\ {{f}^{(0)}}, \ \zeta, \ k_{max}  $\\
		\For{ $k=1:k_{max} $}{
		
	 	$ \bm{s}={{\bm{s}}^{(k-1)}}$
	 	
	 	\text{Compute the matrix} $ \bm{Z}^{(k-1)} $ \text{and vector} $\bm{z}^{(k-1)}$ \text{using} \eqref{equation9} \text{and} \eqref{equation10}
	 	
	 	\text{Compute the matrix }$ {{\bm{\Lambda} }_{1}}^{(k-1)} $ \text{using} \eqref{equation12}
	 	
	 	\text{Update} ${{\bm{h} }^{(k)}}$ \text{using} \eqref{equation14}
	 	
	 	$ \bm{h}={{\bm{h}}^{(k)}} $
	 	
	 	\text{Compute the matrix} $ \bm{Y}^{(k)} $ \text{and vector} $\bm{y}^{(k)}$ \text{using} \eqref{equation17} \text{and} \eqref{equation18}
	 	
	 	\text{Compute the matrix }$ {{\bm{\Lambda} }_{2}}^{(k)} $ \text{using} \eqref{equation20}
	 	
	 	\text{Update} ${{\bm{s} }^{(k)}}$ \text{using} \eqref{equation19}
	 	
 		\text{Update} ${f^{(k)}}$ \text{using} \eqref{equation7}
 		
		\text{Until} $\left| {f^{(k)}} - {f^{(k-1)}} \right| < \zeta $
	}
\end{algorithm}

\section{Experimental Analysis}
To verify the performance of the proposed method, two anti-jamming experiments for the point target scenario and the distributed scenario were carried out, respectively. The transmitted signal is an LFM-PC signal, and the jamming type is repeater deception jamming. The DRFM form of the jammer is established by working in the sample pulse mode. The imaging processing is implemented using RDA with azimuth mismatched filtering, and the constraint weights of the cost function are $\left[ {{\alpha }_{1}},{{\alpha }_{2}},{{\alpha }_{3}},{{\alpha }_{4}} \right]=\left[\text{ 0.2,0.4,0.3,0.1} \right]$, which are applied in both the point-target and the distributed scenarios. This constraint is determined depending on the expected performance of the cost function and should be adjusted in practice when the jamming delay is different. The predefined ISLR is set to ${{R}_{\text{ISL}}}=-\text{60 dB}$, refering to the value of ISRL in \cite{Alighale2014IRSL}. The predefined value of the filtering gain is set to ${{\beta}_{2}}=N$. That is, LPG should be zero in the ideal case. The required SAR parameters are shown below: the carrier frequency is 4 GHz, the pulse width is 5 us, the bandwidth is 200 MHz, and the PRT is 8 ms. Besides, platform information includes the altitude of 2000 m, the aircraft speed of 100 m/s and the radar slant range of 4000 m. Then, we assume the jamming phase delay is 2/3 PRT, the jamming power is 0.04 dB, and the jamming-to-signal ratio (JSR) is $-$0.10 dB.

In the next sections, the effectiveness of the joint waveform modulation and azimuth mismatched filtering algorithm is verified by numerical analysis. Then, the anti-jamming imaging performance of the proposed method is evaluated through imaging simulations of high-fidelity deception jamming scenarios.

\subsection{Convergence Analysis}
\begin{figure}[!t]
	\centering
		\subfigure[]{\label{fig1:(a)}
			\includegraphics[width=1.6 in]{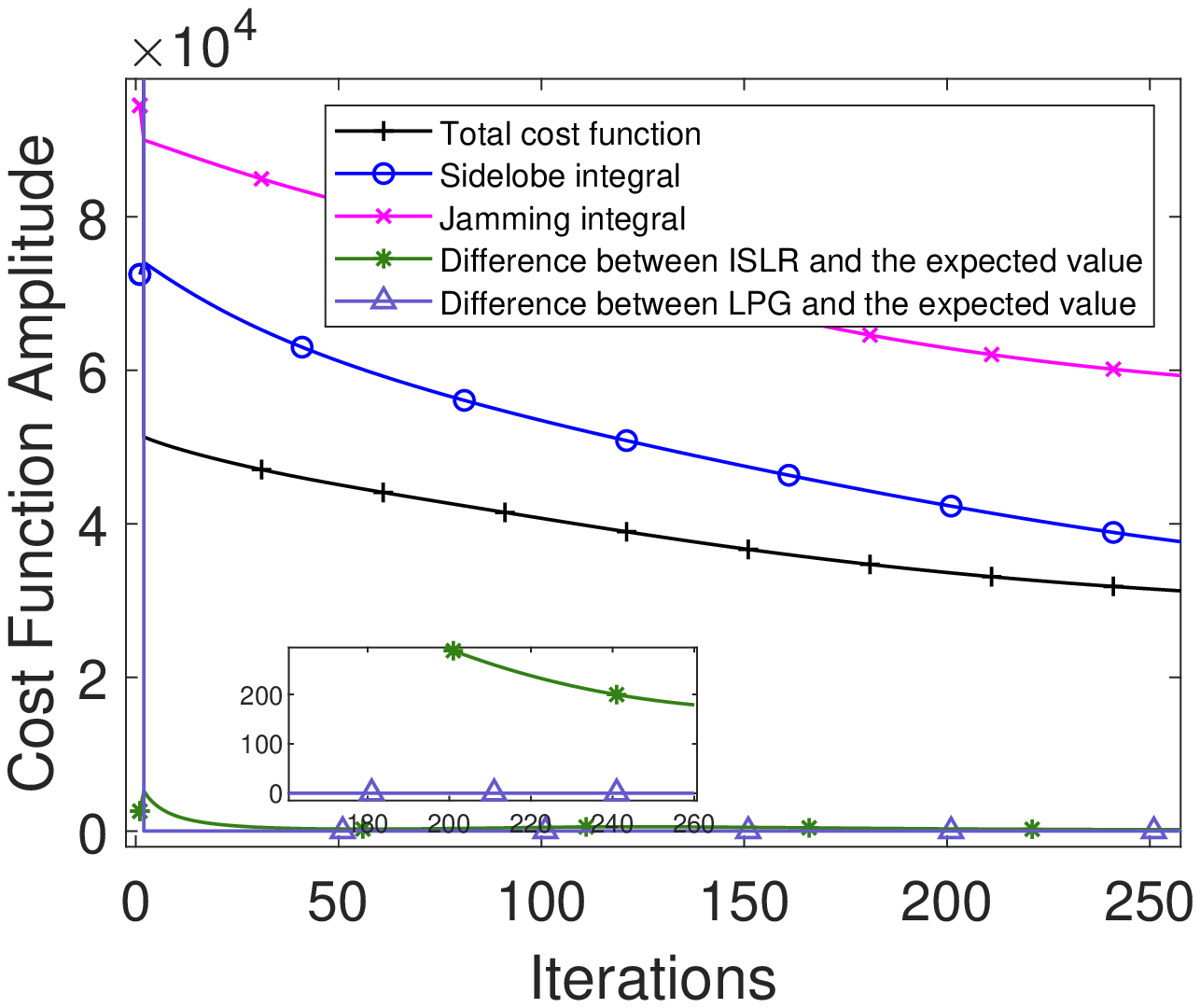}
		}
		\subfigure[]{\label{fig1:(b)}
			\includegraphics[width=1.6 in]{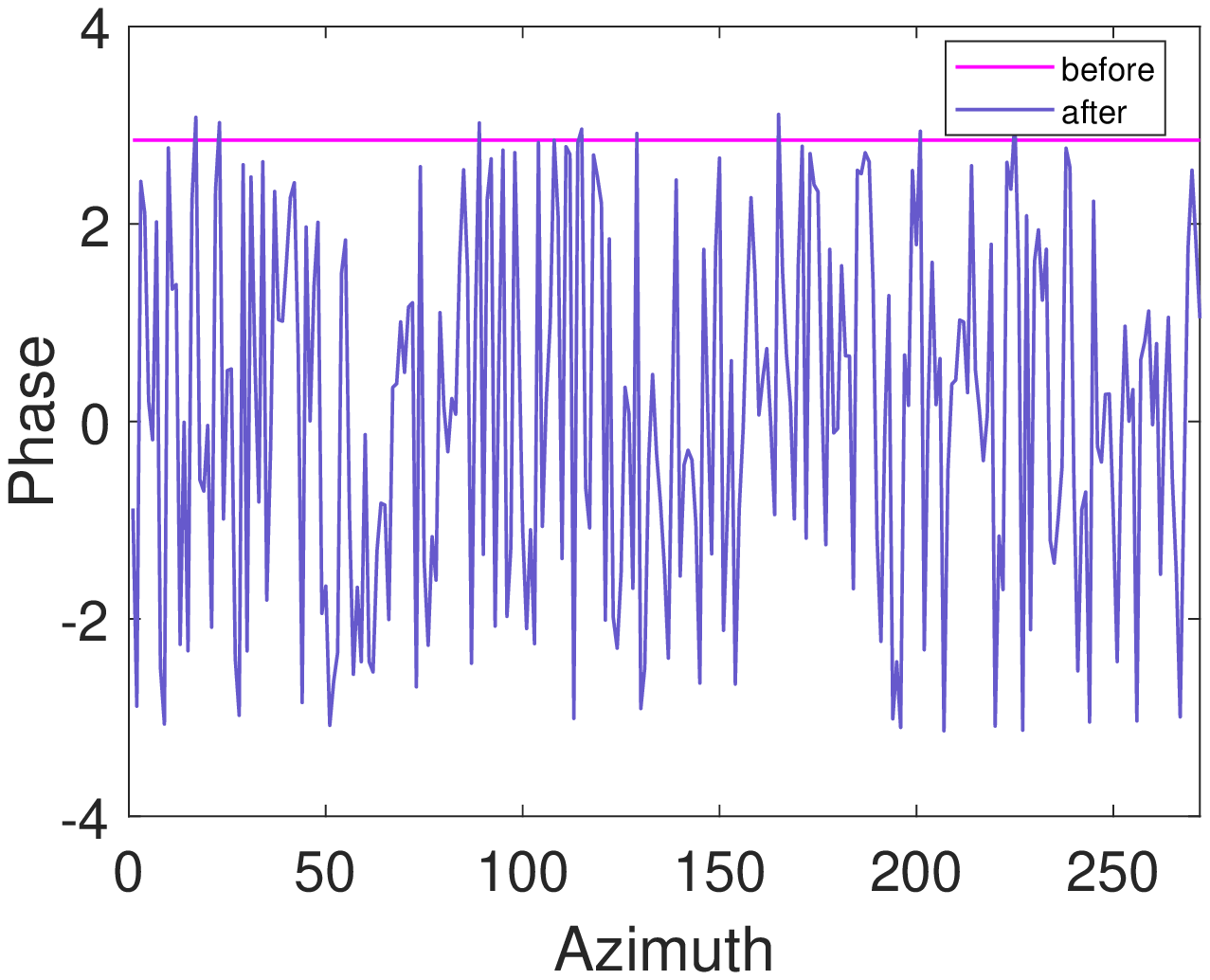}
		}
	\caption{(a) The convergence curves of cost function and its four components; (b) the waveform phase before and after optimization.}
\end{figure}

To prove the validity of the proposed method summarized in Algorithm \ref{alg1}, numerical analysis is carried out. The proposed method adopts the majorization minimization framework and has a non-increasing property. Since the cost function is non-negative, it converges to a certain value with iterations. Additionally, it has also been verified in \cite{Razaviyayn2013convergence} that any sequences generated from the optimization minimization algorithm converge to a stationary point. As shown in Fig. \ref{fig1:(a)}, the curves of the total cost function and the four components of the cost function converge rapidly in the iterative process, indicating that the sidelobe energy of the target signal and the energy of the deception jamming signal are gradually decreasing, and the ISLR and LPG are closer to the preset ideal value, which verifies that the effectiveness of the designed cost function. Fig. \ref{fig1:(b)} shows the phase differential of waveform before and after optimization at a certain fast time. The algorithm realizes phase agility and is conducive to anti-jamming.

\subsection{Point Target Scenario}
\begin{figure}[!t]
	\centering
		\subfigure[]{\label{fig2:(a)}
			\includegraphics[width=1 in]{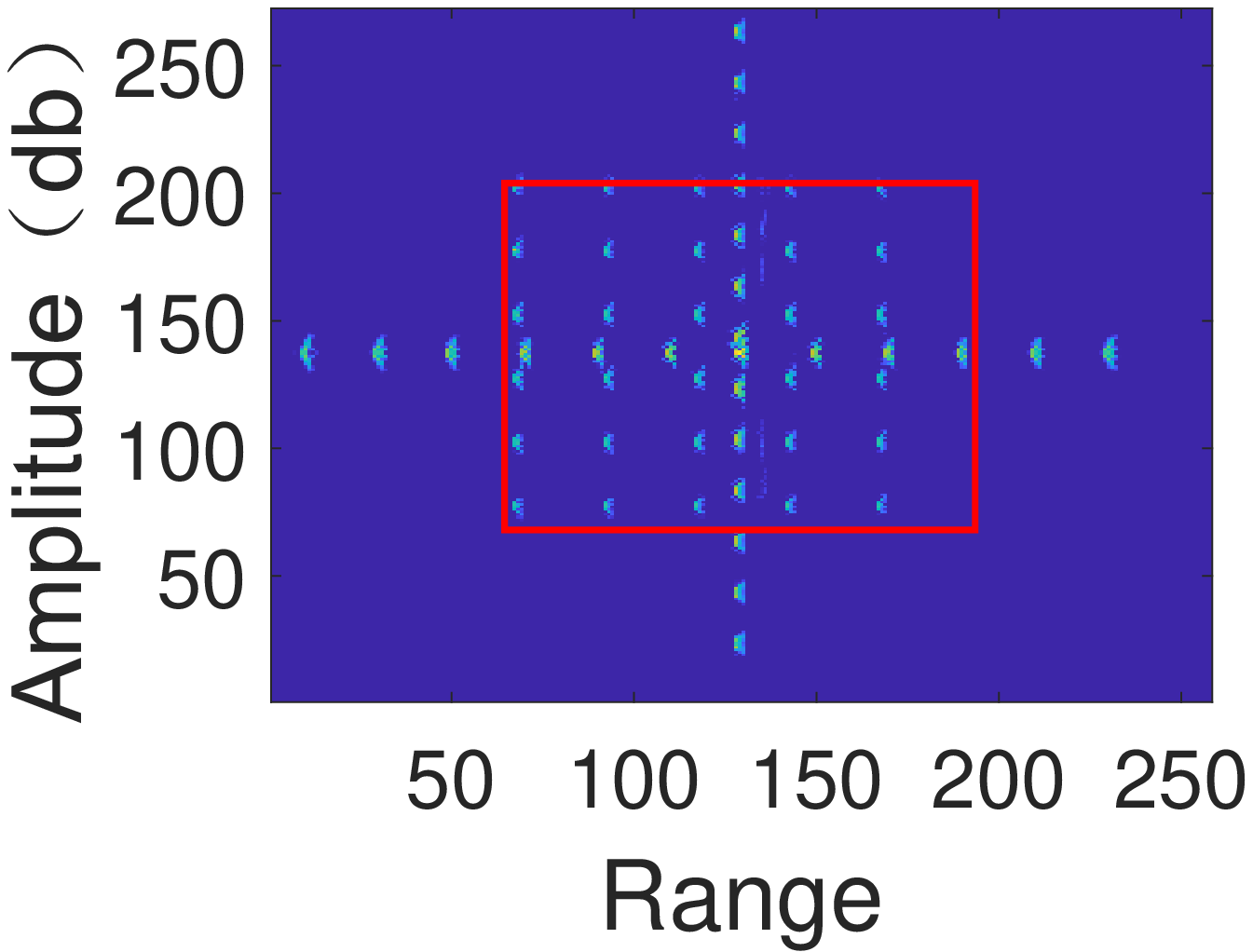}
		}
		\subfigure[]{\label{fig2:(c)}
			\includegraphics[width=1 in]{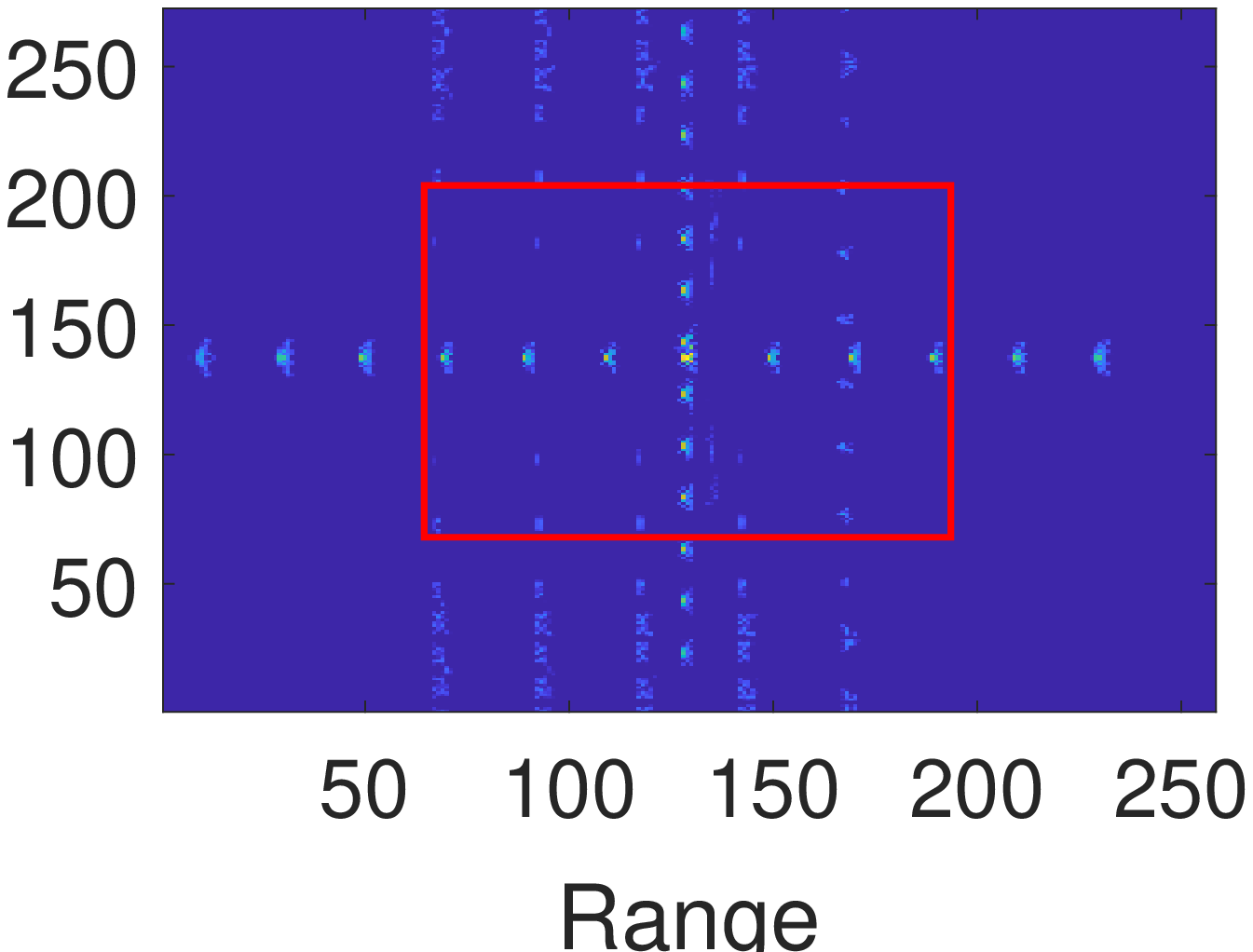}
		}
		\subfigure[]{\label{fig2:(e)}
			\includegraphics[width=1 in]{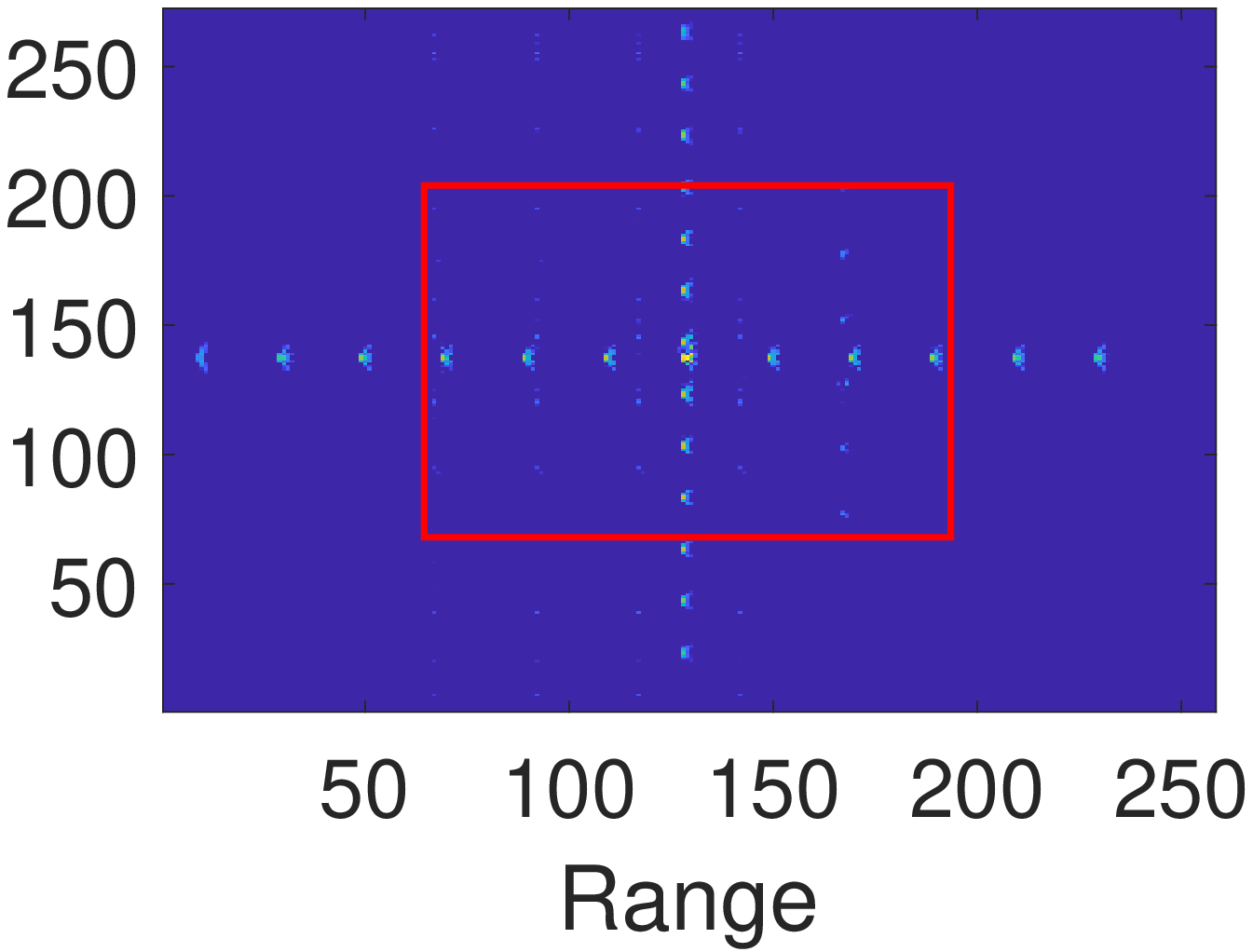}
		}
		\subfigure[]{\label{fig2:(b)}
			\includegraphics[width=1 in]{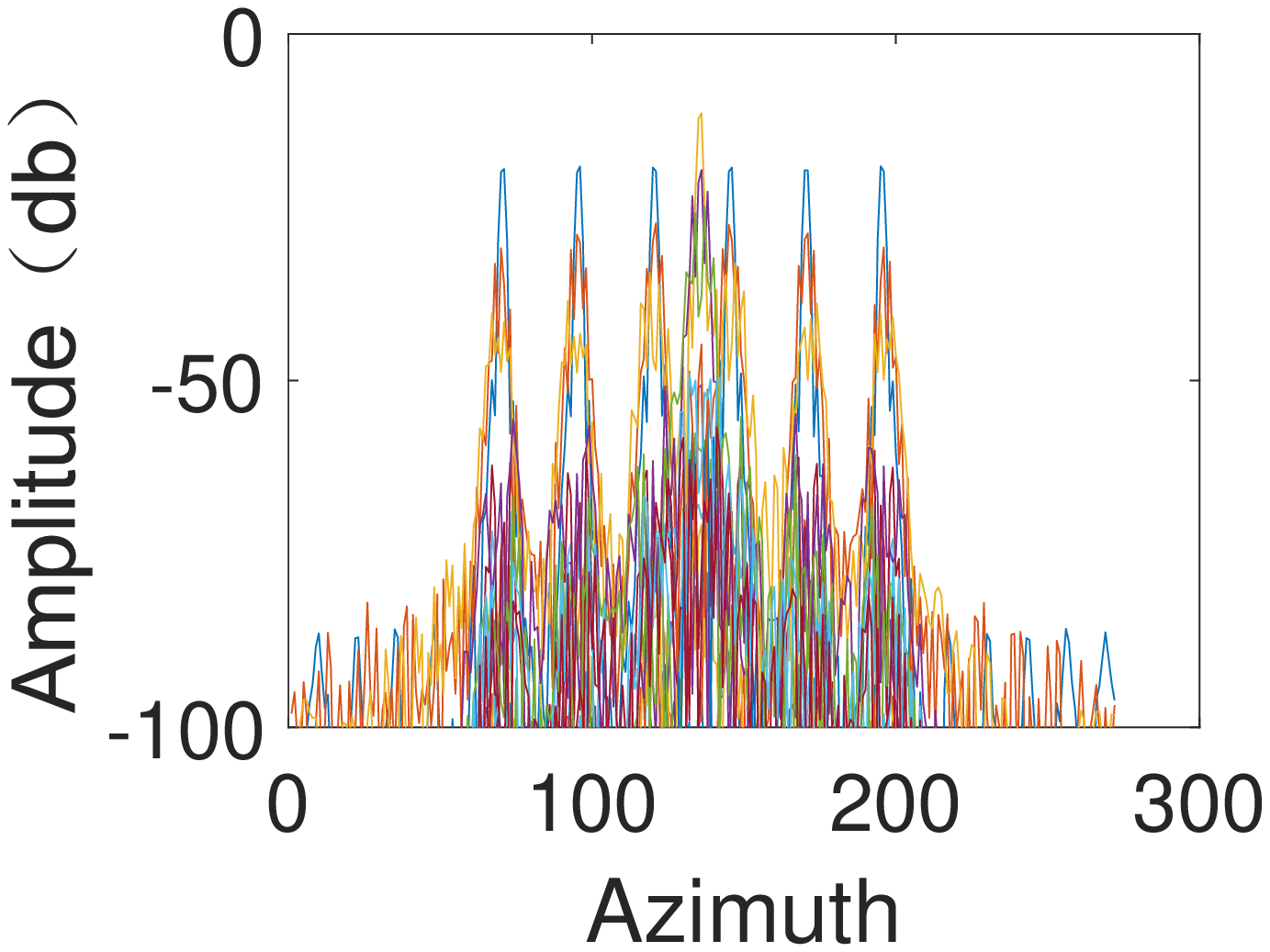}
		}
		\subfigure[]{\label{fig2:(d)}
			\includegraphics[width=1 in]{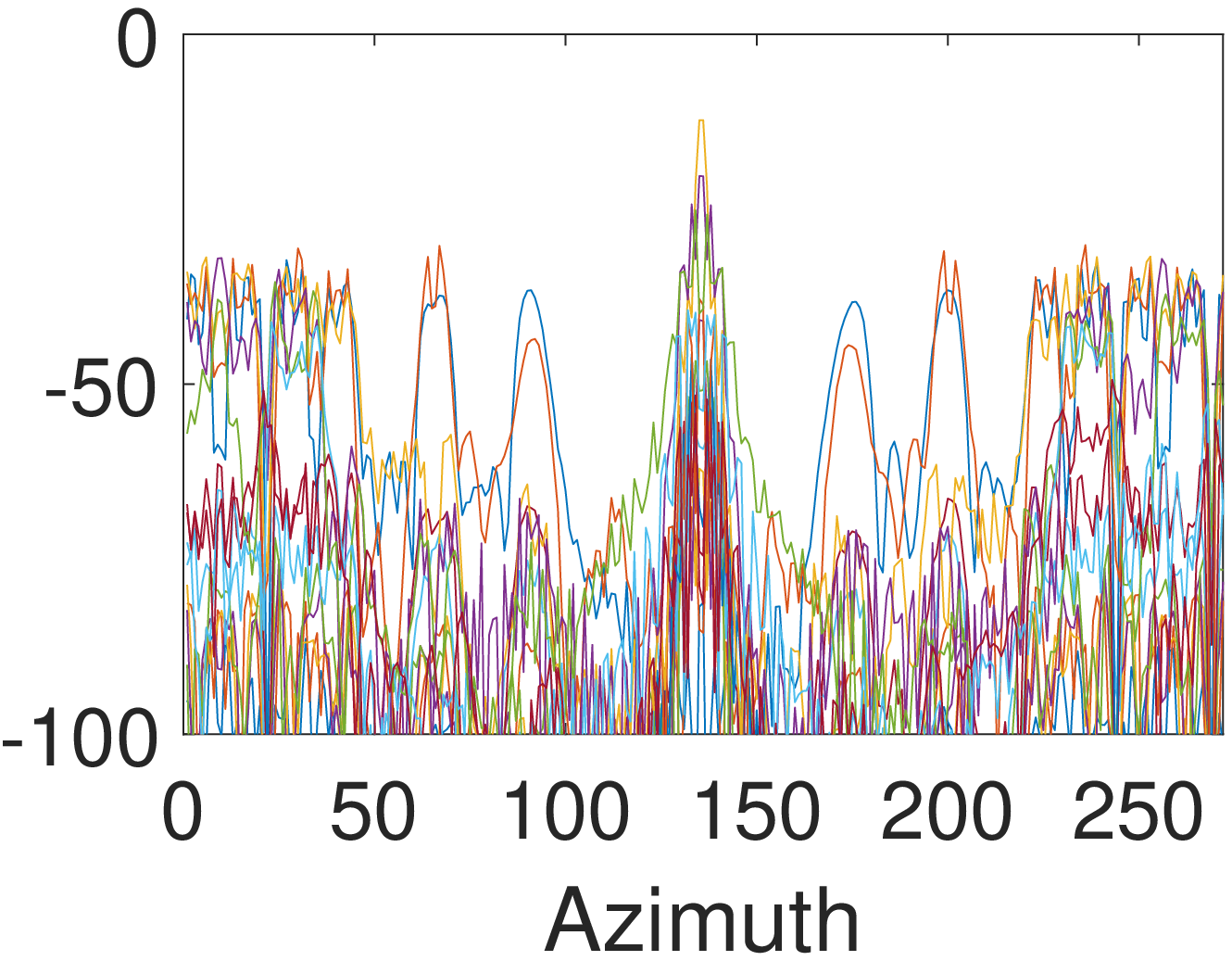}
		}
		\subfigure[]{\label{fig2:(f)}
			\includegraphics[width=1 in]{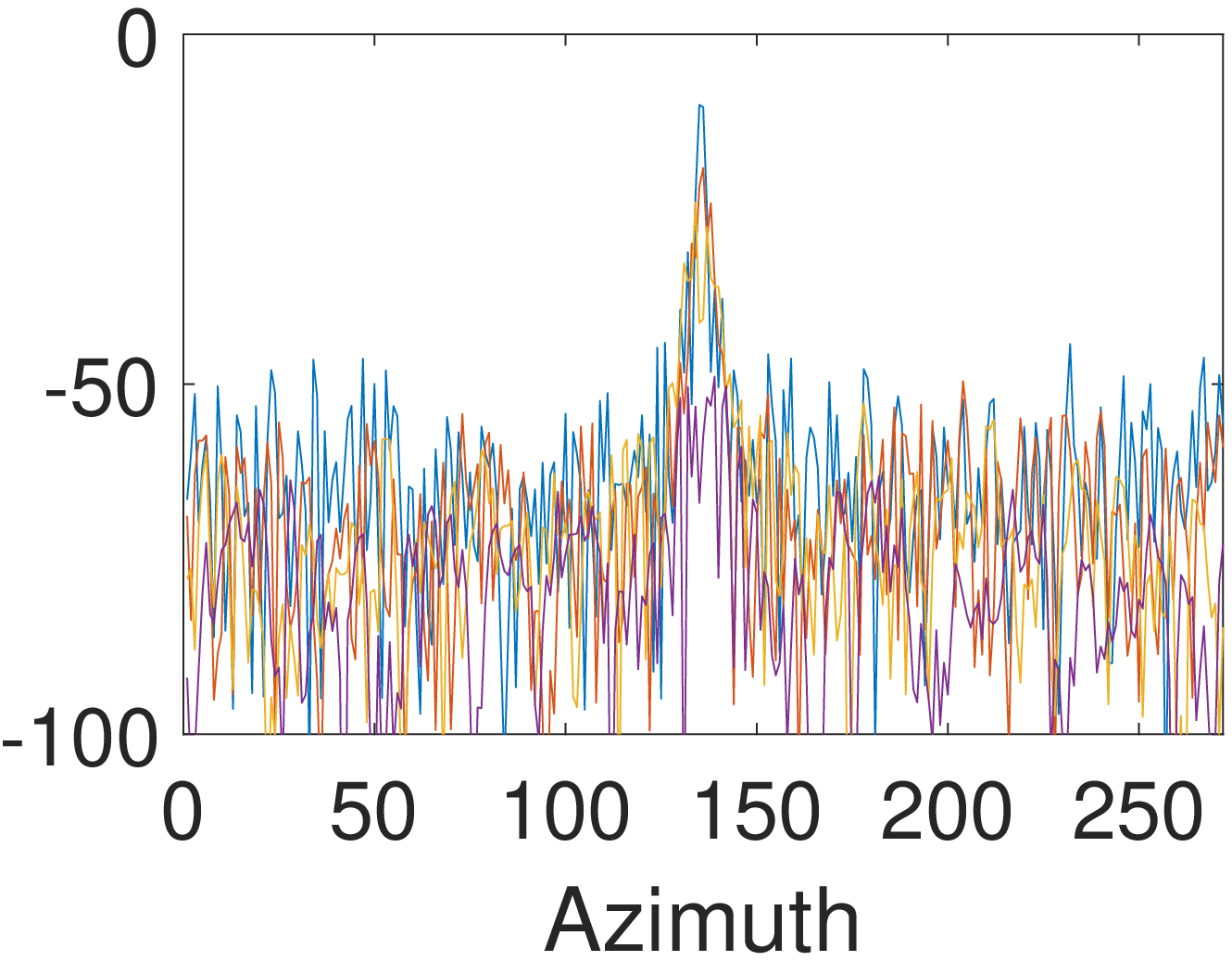}
		}
	\caption{Point target scenario images: (a) jammed image; (b) APC-SAR image (40 dB dynamic range); (c) SAR image of the proposed method (40 dB dynamic range); (d) azimuth profile without anti-jamming; (e) azimuth profile of APC-SAR image; (f) azimuth profile of the proposed method.}
	\label{fig2}
\end{figure}

The point target is centered at the scenario in this experiment, with a cross-shape distribution of 20 m intervals in azimuth and range directions, respectively. The jammed point target is distributed in a square shape with the central coordinate $(0,0)$, which is marked using the red box in Fig. \ref{fig2:(a)}. In this section, the proposed method is compared with the azimuth phase coding (APC) method \cite{Tang2021HighFidelity} of anti-jamming in the azimuth direction. Without any anti-jamming methods, the SAR imaging performance of a point scenario is shown in Fig. \ref{fig2:(a)} and Fig. \ref{fig2:(b)}. When the APC method is adopted, the SAR imaging performance of a point scenario is shown in Fig. \ref{fig2:(c)} and Fig. \ref{fig2:(d)}. Using the proposed method in this paper, the SAR imaging performance of point scenario is shown in Fig. \ref{fig2:(e)} and Fig. \ref{fig2:(f)}. Fig. \ref{fig2:(b)}, Fig. \ref{fig2:(d)} and Fig. \ref{fig2:(f)} show the azimuth amplitude of the same disturbed range, where the central value is the target signal response, and the rest are the jamming signal response.

When the reserved energy ranges from $-$40 dB to 0 dB,  both the APC method and the proposed method can effectively achieve the same performance, according to the comparison of Fig. \ref{fig2:(a)}, Fig. \ref{fig2:(c)} and Fig. \ref{fig2:(e)}. The real point target is clearly visible and the jamming signal is filtered out. The APC method causes the jamming signal energy to shift to both sides of the azimuth direction, bringing deception jamming to other areas, as shown in Fig. \ref{fig2:(b)}, Fig. \ref{fig2:(d)} and Fig. \ref{fig2:(f)}. The proposed method makes the jamming signal discrete upward in the whole azimuth direction, which can efficiently reduce the intensity of the jamming signal. In addition, due to the fixed mechanism of the APC method, the modulated signal still has the risk of being intercepted and counterattacked. In contrast, the designed parameters of the proposed method are optimized using the characteristics of the jammer, which has certain adaptive ability in practice ant-jamming imaging.

In order to verify the imaging performance of the proposed method, structural similarity (SSIM), mutual information (MI), and JSR are introduced as the evaluation criteria, where SSIM is defined as formula (40) in \cite{Zhou2022Waveform}, MI is defined as formula (6) in \cite{Gao2008MI} and JSR is defined as formula (25) in \cite{Li2020JSR}. The specifics are shown in Table \ref{tab:table2}. SSIM represents the retention degree of the target signal. As shown Table \ref{tab:table2}, the proposed method reduces the jamming intensity while better retaining the target signal.

\begin{table}[!t]
	\caption{Evaluation Index Value}\label{tab:table2}
	\centering
	\begin{tabular}{cccc }
		\toprule
		method & SSIM & MI & JSR\\
		\midrule
		Without anti-jamming  & 1.00  & 0.81 & $-$5.95 dB  \\
		APC   & 0.99 &  0.83 & $-$8.84 dB\\
		The proposed method  & 0.99 &  0.83 & $-$12.01 dB\\
		\bottomrule
	\end{tabular}
\end{table}
\begin{figure}[!t]
	\centering
		\subfigure[]{\label{fig3:(a)}
			\includegraphics[width=1.6in]{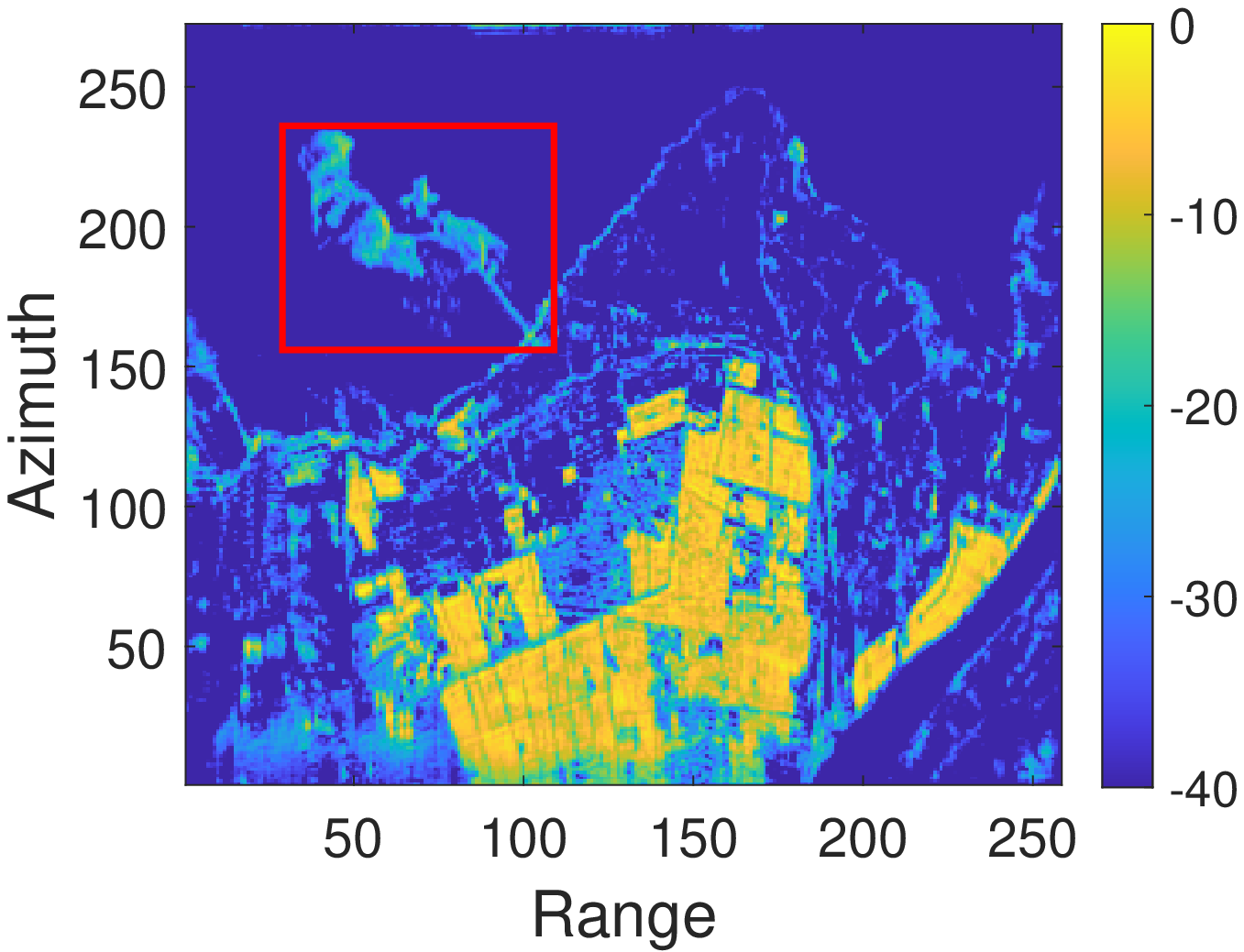}
		}
		\subfigure[]{\label{fig3:(b)}
			\includegraphics[width=1.6in]{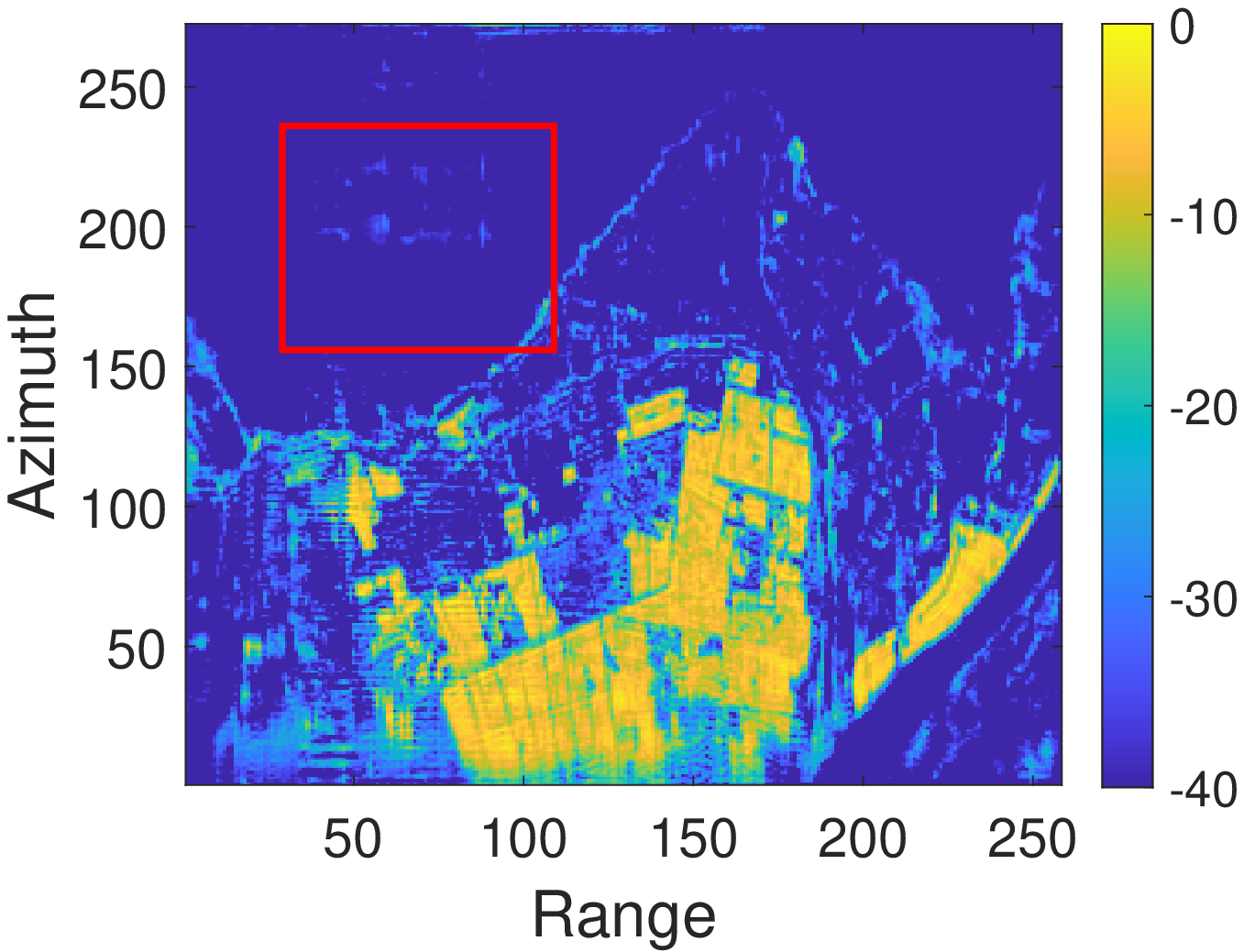}
		}
		\subfigure[]{\label{fig3:(c)}
			\includegraphics[width=1.6in]{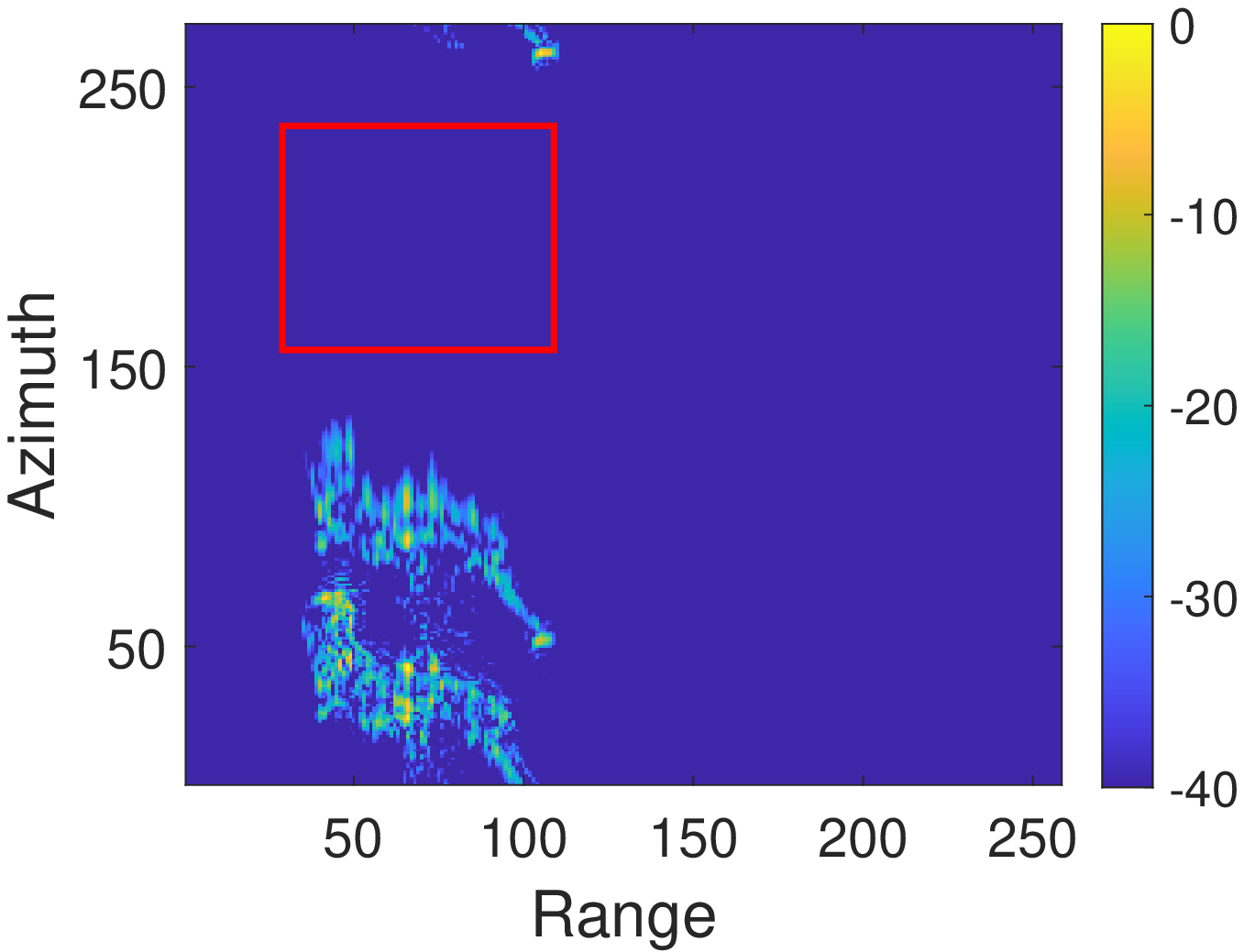}
		}
		\subfigure[]{\label{fig3:(d)}
			\includegraphics[width=1.6in]{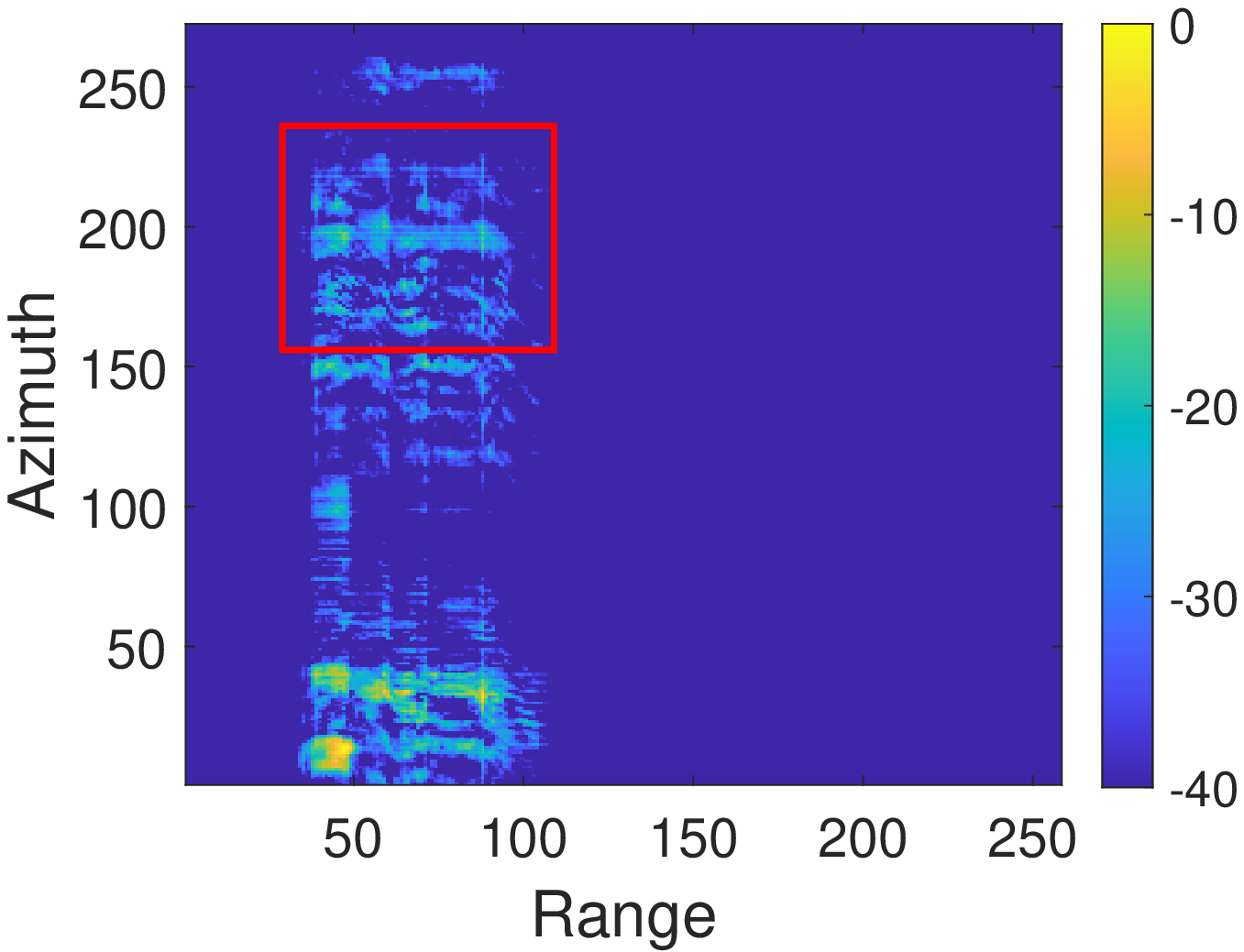}
		}		
	\caption{Distributed scenario images: (a) jammed image; (b) image of the proposed method; (c) APC-SAR jammed areas image; (d) SAR jammed areas image of the proposed method.}
	\label{fig3}
\end{figure}
\subsection{Distributed Scenario}
In this case, the experimental scenario was selected from the Hongze Lake, Liuzi River basin in China. The jamming signal comes from a faked port target. The central point of the jammed scenario is located at ($-$60 m, $-$60 m). Without any anti-jamming processing, the imaging quality is shown in Fig. \ref{fig3:(a)}, where the deception scenario is clearly visible on the image. The anti-jamming imaging performance with the proposed method is shown in Fig. \ref{fig3:(b)}. The deception scenario is effectively filtered out, and the residual jamming level is less than $-$40 dB. The processing performance of the APC algorithm and the proposed algorithm on deception is shown in Fig. \ref{fig3:(c)} and Fig. \ref{fig3:(d)}. After anti-jamming using the proposed method, the energy of the jamming signal is dispersed in the whole azimuth dimension, reducing the peak energy of deception jamming, while the APC algorithm only shifts it.

\section{Conclusion}
In this paper, a joint design method of waveform modulation and azimuth mismatched filtering was proposed. The multi-objective optimization function is constructed to suppress the jamming level and keep the real target signal, according to the evaluation criterion of the SAR image. An optimization algorithm with low computational complexity was proposed to estimate the initial phase and azimuth mismatched filter coefficients. Through the convergence analysis of the objective function and the anti-jamming experiments of the point target scenario and distributed scenario, the effectiveness of the proposed method against deception jamming was verified. The joint waveform and filter design method was designed based on the prior knowledge that the jammer has a fixed time delay. The anti-jamming performance is degraded if the prior delay knowledge of the jammer deviates greatly from the actual delay of the jamming system. In practice, it can be optimized using adaptive adjustment. We can iteratively estimate the exact time delay of the jammer through the analysis of jamming effect feedback to ensure the universality of the proposed method.

\ifCLASSOPTIONcaptionsoff
\newpage
\fi

\bibliographystyle{IEEEtran}
\bibliography{IEEEabrv,myIEEEbibfile}

\end{document}